# Improving Snore Detection under Limited Dataset through Harmonic/Percussive Source Separation and Convolutional Neural Networks*,**

F.D. González-Martínez[a,∗,1], J.J. Carabias-Orti[1], F.J. Cañadas-Quesada[1], N. Ruiz-Reyes[1], D. Martínez-Muñoz[1] and S. García-Galán[1]

[a]Departament of Telecommunication Engineering. University of Jaen, Campus Científico-Tecnologico de Linares, Avda. de la Universidad, s/n, Linares (Jaen), 23700, Spain

## ARTICLE INFO

## ABSTRACT

Snoring, an acoustic biomarker commonly observed in individuals with Obstructive Sleep Apnoea Syndrome (OSAS), holds significant potential for diagnosing and monitoring this recognized clinical disorder. Irrespective of snoring types, most snoring instances exhibit identifiable harmonic patterns manifested through distinctive energy distributions over time. In this work, we propose a novel method to differentiate monaural snoring from non-snoring sounds by analyzing the harmonic content of the input sound using harmonic/percussive sound source separation (HPSS). The resulting feature, based on the harmonic spectrogram from HPSS, is employed as input data for conventional neural network architectures, aiming to enhance snoring detection performance even under a limited data learning framework. To evaluate the performance of our proposal, we studied two different scenarios: 1) using a large dataset of snoring and interfering sounds, and 2) using a reduced training set composed of around 1% of the data material. In the former scenario, the proposed HPSS-based feature provides competitive results compared to other input features from the literature. However, the key advantage of the proposed method lies in the superior performance of the harmonic spectrogram derived from HPSS in a limited data learning context. In this particular scenario, using the proposed harmonic feature significantly enhances the performance of all the studied architectures in comparison to the classical input features documented in the existing literature. This finding clearly demonstrates that incorporating harmonic content enables more reliable learning of the essential time-frequency characteristics that are prevalent in most snoring sounds, even in scenarios where the amount of training data is limited.

## 1. Introduction

Obstructive Sleep Apnoea Syndrome (OSAS), as defined by the World Health Organization (WHO), is a clinical disorder characterized by recurring episodes of nocturnal apnoeic events, often accompanied by prominent snoring. These episodes result in temporary interruptions of airflow, leading to intermittent hypoxemia because of upper airway collapse [1]. Moreover, the repeated episodes during the night make it difficult to achieve a restful sleep, so it implies a range of adverse psychological states, such as continuous fatigue, irritability and depressive symptoms [2]. As previously mentioned, snoring is the most common feature exhibited by individuals with OSAS during the night [1, 3]. Although OSAS affects approximately $3 - 7\%$ of adult men and $2 - 5\%$ of adult women worldwide with observed correlations to advancing age, body weight and alcohol consumption [4, 5], snoring sounds affects 44% of the male population while it affects only 28% of the female population, both between 30 and 60 years of age [6]. Several studies have demonstrated that snoring sounds can provide useful information in detecting and assessing OSAS [7, 8, 9]. Because OSAS is costly due to its widespread occurrence, the signal processing and artificial intelligence (AI) research communities are dedicating significant resources to developing more advanced systems based on snoring sounds for early detection and assessment of OSAS severity [10, 11, 12]. These systems aim to use medical information from contactless methods based on snoring sounds instead of relying on the contact methods such as, the current gold standard device full-night polysomnography (PSG), which requires extensive human resources and may cause patient discomfort.

Snoring is an audible respiratory sound that predominantly manifests during the inspiratory phase of the respiratory cycle, although it may also occur during the expiratory phase [13]. The underlying cause of snoring is a decrease in muscular tone of the upper airway during sleep, resulting in narrowing and increased resistance at this level that leads to partial obstruction. Consequently, turbulent airflow and vibrations of the pharyngeal soft tissues occur as air passes through [7]. Despite the absence of a precise definition of snoring based on specific acoustic parameters [7], a snoring event typically has a duration ranging from 300 ms to 3 s, with an average duration of approximately 1.5 s [14, 15]. Furthermore, snoring generally encompasses a frequency spectrum that typically extends up to 4000 Hz [7, 16, 17, 18], with a predominant concentration of acoustic energy between 50 and 250 Hz [19, 20, 21]. Focusing on the anatomical structures that contribute to airway obstruction, it is widely known the VOTE classification scheme [22] to categorize four types of snoring sounds: a) Velum (V), that includes the soft palate, uvula, and lateral pharyngeal wall tissue at the level of the velopharynx; b) Oropharyngeal lateral walls (O), which encompass the palatine tonsils and the lateral pharyngeal wall tissues; c) Tongue base (T) and finally, d) Epiglottis (E). Common to each category is the presence of snoring sounds that show harmonic behaviour observed through narrowband energy trajectories over time, as can be observed in Figure 1. However, this harmonic nature is predominantly observed in V- and O-type snoring, while it is comparatively less prominent in T- and E-type snoring.

*This work was supported in part under grant PID2020-119082RB-C21 funded by MCIN/AEI/10.13039/501100011033 and grant P18-RT-1994 funded by the Ministry of Economy, Knowledge and University, Junta de Andalucía, Spain.

∗Corresponding author
✉ fdgonzal@ujaen.es (F.D. González-Martínez)
ORCID(s): 0009-0008-0667-8513 (F.D. González-Martínez)

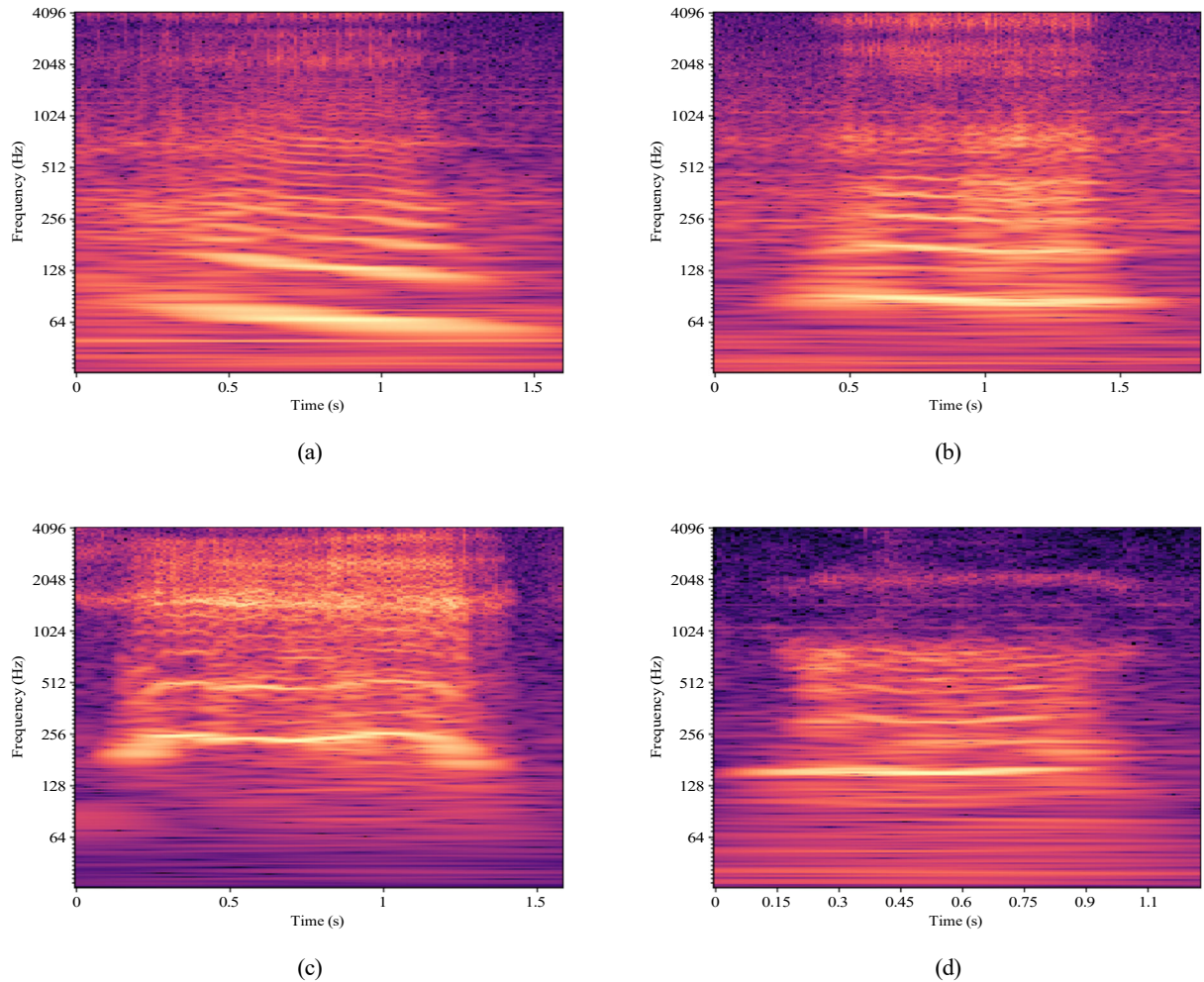

**Figure 1:** Time-frequency representation (spectrogram) for some snoring events of classes V (a), O (b), T (c) and E (d).

Over the past decades, numerous studies have been conducted to investigate snore detection from respiratory sounds. Most of these snore detection methods are based on different signal processing and AI techniques such as, Hidden Markov Models (HMM) [23], subband energy [24], fuzzy clustering [25], AdaBoost classifier [26], random forest [27], k-Nearest Neighbor (KNN) [28], logistic regression [29], Gaussian Mixture Models (GMM) [30], Gray Wolf Optimization (GWO) [31], Linear Predict Coding (LPC) [32], Support Vector Machine (SVM) [33, 34] and local binary patterns [35]. Duckitt et al. [23] utilized HMMs to differentiate snoring from ambient sounds based on Mel-Frequency Cepstral Coefficients (MFCCs). A snore sound detection system is proposed [26] employing an AdaBoost classifier, which was supplied with 34 discriminative time-frequency features extracted from manually labelled acoustic events. Nonaka et al. [29] developed a novel method for automatically extracting snore sounds from sleep sounds by using an auditory image model (AIM). The AIM simulates the auditory pathway from the outer ear to the cerebrum, and it was employed to obtain the features to perform the snore/non-snore classification through a multi-nominal logistic regression (MLR) analysis. Wang et al. [33] proposed a method for identifying snoring event by means of acoustic feature analysis with an SVM. In [30], an unsupervised segmentation and classification of snoring events is developed based on statistical analysis through the Expectation-Maximization algorithm applied to Gaussian Mixture Models to cluster the data set according to the probability of each data point of belonging to some normal distribution component. Recently, deep learning (DL) has emerged as the standard technique for snoring detection, demonstrating excellent results when there is a substantial amount of data available using different approaches such as Recurrent Neural Networks (RNN) [36, 37], Convolutional Neural Network (CNN) [38, 39, 40, 41], Feed-forward Artificial Neural Networks (FANN) [42] and Convolutional and Recurrent Neural Network (CRNN) [15]. In this DL context, Lim et al. [37] introduced an RNN approach in which the zero-crossing rate (ZCR), Short Time Fourier Transform (STFT) and MFCCs were employed for feature extraction. Jiang et al. [39] presented an automatic snoring detection algorithm that utilizes CNN descriptors extracted from audio maps from 15 patients, and their findings revealed that the highest performance metrics were achieved by a model comprising CNNs, long short-term memory networks (LSTMs), and deep neural networks (DNNs) fed with the Mel-spectrogram. Xie et al. [15] developed a snore detection algorithm by combining a CNN and an RNN, where the last one processed the sequential CNN output for classifying audio events into snore and non-snore categories. The results achieved an accuracy of 95.3%, a sensitivity of 92.2%, and a specificity of 97.7%, confirming that the microphone positioned at a distance of approximately 70 cm above the subject's head achieved the highest accuracy. However, to the best of our knowledge, previous studies have not focused on performance when available data are scarce when feeding the DL architecture applied to snoring detection. This aspect is particularly significant in the case of snoring data due to the inherent challenges associated with its acquisition. These challenges include the requirement for multiple audio recordings in a hospital sleep room and the significant human resources needed for labelling the audio segments.

The fidelity of snore detection is closely linked to the dependability of the extracted feature set utilized for snoring sound recognition. Nevertheless, the acoustic content of these features may be undermined when snoring events exhibit overlapping occurrences with high-energy noise. Consequently, techniques based on noise reduction (denoising) and source separation become valuable tools to be applied in real-world environments within the detection model [12]. Most of the denoising methods are based on least-Mean-Square adaptive filter [43], short-term energy [29], Wiener filtering [44], spectral subtraction [33], average amplitude thresholding [14, 36], bandpass filtering [15] and wavelet [45, 42]. On the other hand, most of the source separation approaches propose Blind Delayed Source Separation (BDSS) [46], Non-negative Matrix Factorization (NMF) [47, 45], Principal and Independent Component Analysis (PCA/ICA) [48] and Multivariate Variational Mode Decomposition (MVMD) [45]. However, none of them are particularly designed to extract most of the harmonic content shown by snoring sounds. For this reason, our proposal is focused on detecting monaural snoring sounds from non-snoring sounds extracting the harmonic content of the input sound by means of harmonic/percussive sound source separation (HPSS), a technique widely used in music signal processing [49, 50, 51, 52], and more recently, in the analysis of biomedical adventitious sounds [53]. The main contribution of this study improves the learning process for deep learning architectures in the domain of snore detection. This improvement is achieved by leveraging the predominant harmonic content extracted from snoring sounds as a novel feature, which is subsequently incorporated into conventional state-of-the-art CNN models, being particular promising within the context of the limited data learning paradigm. Our motivation comes from an empirical observation that a significant proportion of snoring sounds exhibit varying degrees of harmonic behaviour. This insight suggests that leveraging such content could be beneficial, especially in contexts where training data availability is limited.

The paper is organized as follows: Section 2 describes the database. Section 3 details the proposed HPSS-based method. Section 4 focuses on the evaluation setup, including the metrics used and the parameters of the neural network architectures, together with the snore detection results evaluating several time-frequency representations and training sizes. Section 5 provides an analysis of the behavior derived from the most relevant performance results of the proposed method. Finally, Section 6 summarizes the conclusions derived from the study and outlines directions for future research.

## 2. Materials and methods

To evaluate the proposed snoring detection method, a $D_T$ database composed of snoring monaural sounds from the $D_S$ database and non-snoring monaural sounds from the $D_N$ database has been created. In this work, snoring sounds are mixed with non-snoring sounds (clinical ambient sounds, household noises, room sounds, and cough sounds) that have a high probability of occurring in real acoustic environments. However, both snoring and non-snoring sounds have not been captured by the authors but have been collected from previous referenced works and repositories widely used in this biomedical field where spatial sound information is not available. These previous databases and snoring and non-snoring sounds are explained in detail below.

The snoring sounds dataset $D_S$ encompasses 2500 monaural snoring events from 50 participants [42] and 828 monaural snoring events from 219 patients [14]. Consequently, $D_S$ is composed of 3328 snoring sound events belonging to 269 subjects. As CNN architectures require input samples of the same size, a histogram has been applied to determine the minimum temporal duration to ensure that over 97% of the snoring sounds exhibit the required temporal duration or less. As a result, the duration of each snoring sound has been set to 3.5 s, similar as occurs in [15]. Hence, each snoring sound event with a duration less than 3.5 s has been padded with zeros to match the specified temporal duration, while each snoring sound event with a duration exceeding 3.5 s has been truncated.

| Type | Total number of sounds | Mixed ($D_S + D_N$) | | Unmixed ($D_N$) | |
|---|---|---|---|---|---|
| | | Training (90%) | Test (10%) | Training (90%) | Test (10%) |
| Clinical [54] | 2500 | 749 | 84 | 1500 | 167 |
| Household [58, 59] | 2500 | 749 | 83 | 1500 | 168 |
| Cough [63] | 2500 | 749 | 83 | 1500 | 168 |
| Room [42] | 2500 | 748 | 83 | 1501 | 168 |

**Table 1**
A detailed overview of the training and test sets used in database $D_T$. The term "Mixed ($D_S + D_N$)" represents a sound mixture composed of a snore sound from $D_S$ and an interfering sound from $D_N$. The term "Unmixed ($D_N$)" represents an interfering sound from $D_N$. The set of interfering sounds used in "Mixed" mixtures were distinct from those in "Unmixed".

The non-snoring sounds dataset $D_N$ was initially composed of 10000 monaural audio events categorized in four classes of non-snoring sounds: a) Clinical ambient sounds [54], which are widely used in biomedical signal processing [55, 56, 57]. These sounds have been extracted from 75 audio files of variable duration that were recorded in different real clinical environments. To maintain concordance with the previously defined snoring events, these audio files were partitioned into 3.5 s segments, producing a total of 2744 clinical noise events, of which only 2500 were randomly chosen for inclusion in the dataset; b) Household noises from the DESED dataset [58, 59], which is well known in the field of sound event detection [60, 61, 62]. A subset of 2500 sound events recorded in a real (non-synthesized) home environment were randomly selected; c) Room sounds [42], composed of 2500 events characterized by shallow breaths or ambient quietness commonly observed in a clinical or home room; d) Cough sounds [63]. A subset of 2500 sound events composed of cough. Subsequently, a total of 3328 non-snoring events were extracted from $D_N$ following an equal distribution among the four classes mentioned above to mix them with the snoring events, while the remaining 6672 non-snoring events were used to form the set of interfering non-snoring events $D_N$. Hereafter, an interfering sound refers to a non-snoring sound.

To enhance the snoring detection system's robustness in scenarios where snoring sounds overlap with non-snoring sounds, we introduced modifications to the database $D_S$. Specifically, each snoring sound from $D_S$ was combined with a non-snoring sound from $D_N$ using different signal-to-noise ratios (SNRs) of −5 dB (referred to as database $D_{T_{-5}}$), 0 dB (referred to as database $D_{T_0}$), and 5 dB (referred to as database $D_{T_5}$). As a result, three noisy snoring signals, taking into account the aforementioned SNR values, have been generated for each snoring event, resulting in a total of 9984 noisy snoring events. Finally, the events of each sound class have been divided into training and test sets, with 90% for training and 10% for testing, ensuring subject independence since data from the same subject are not used in both the training and test sets simultaneously. The distribution of the database $D_T$, for an arbitrary SNR, is described in Table 1.

## 3. Proposed method

Although most conventional feature extraction approaches, shown in Figure 2, are based on different time-frequency representations, such as Short-Time Fourier Transform (STFT), Mel-spectrogram and Constant-Q Transform (CQT), the proposed method is composed of three stages: (i) Preprocessing (based on CQT), (ii) Feature extraction based on HPSS and (iii) Neural network architecture. This method is detailed through this section and its flowchart is depicted in Figure 2.

### 3.1. Stage 1. Preprocessing

In the following subsection, we detail the parameters used to obtain the different time-frequency representations that will feed the neural network architectures used in this work.

- Short-Time Fourier Transform (STFT). For its calculation, the database $D_T$ was resampled to 8 kHz as proposed in [64] since the predominant energy of snoring is below 4 kHz. Moreover, a Hanning window of 256 samples and a hop size of 64 samples were utilized. Note that STFT may not be the best choice for analysing respiratory sounds because this time-frequency representation yields a linear frequency scale, resulting in decreased resolution at lower frequencies where most of the relevant respiratory spectral content such as, snoring sounds exists. As an example, STFT has been observed to exhibit subpar performance when analysing auscultated sounds in noisy environments, as reported in [65].

- Mel-spectrogram. The Mel-scaled spectrogram is based on the short-term energy spectrum at a logarithmic frequency scale that corresponds to the human ear perception [66]. In this case, 32 Mel filters were applied to the previously obtained STFT, since it has been demonstrated that a similar number of Mel filters provides good resolution in the frequency bands for classifying respiratory sounds [67].

- Constant-Q Transform (CQT). The CQT is an extension of the STFT using a time-varying window in order to compute the logarithmic frequency spectrum so that the center frequencies of the frequency bins are geometrically spaced and their Q-factors are all equal.

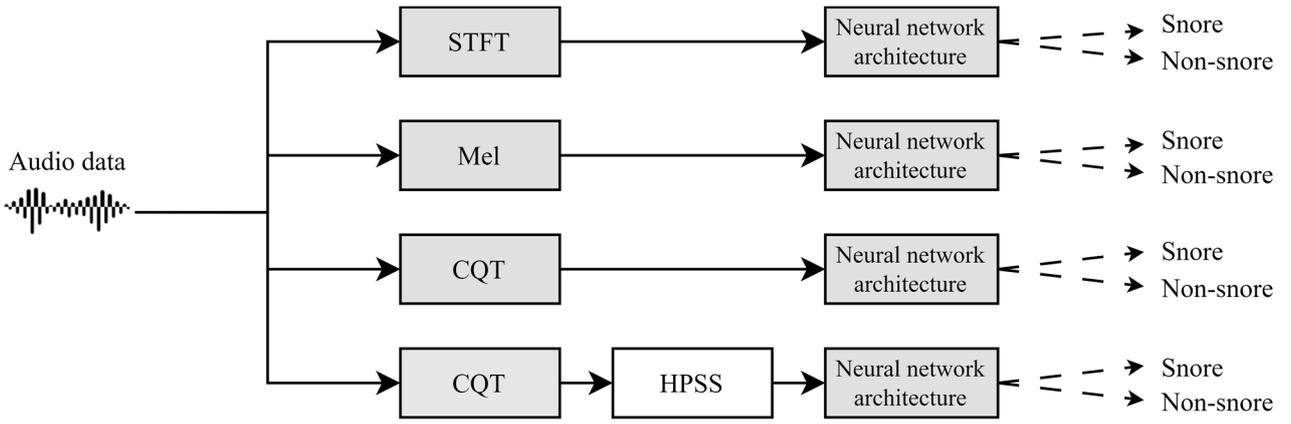

**Figure 2:** Flowchart of conventional feature extraction based on STFT, Mel and CQT (three upper branches). In the flowchart of the proposed method (lower branch), the feature extraction is based on the incorporation of HPSS at the output of the CQT to feed the neural network more efficiently.

As a consequence, frequency resolution is better for low frequencies and time resolution is better for high frequencies. In this work we have used the implementation by [68]. In contrast to the STFT, the database $D_T$ was resampled to 48 kHz in order to replicate the same parameters proposed in [15], specifically, a Hanning window with a hop length of 512 samples and a bin number of 84, with 12 bins per octave. Focusing on the feature extraction based on HPSS explained in the following section, we have applied the same parameter configuration to obtain the CQT spectrogram.

Subsequently, all previously computed complex spectrograms were converted to magnitude spectrograms in order to compress the input data for the various neural network architectures.

### 3.2. Stage 2. Feature extraction based on HPSS

The harmonic spectrogram extracted by HPSS is used as a feature to feed input data into conventional neural network architectures in order to improve the snoring detection performance. The motivation is that snoring sounds commonly show a remarkable harmonic structure, as previously mentioned. Consequently, leveraging this particular content could be useful in acoustic scenarios where training data is limited. In this work, HPSS has been implemented by using the method described in [50]. In particular, HPSS applies median filtering across time to attenuate percussive events and accentuate the harmonic components. On the other hand, HPSS uses median filtering across frequency to remove harmonic events and enhance the percussive components. Therefore, the harmonic filtered spectrogram $H$ and the percussive filtered spectrogram $P$ can be obtained as follows,

$$H(k, m) = \{X(k, m - \lfloor l_h/2 \rfloor : m + \lfloor l_h/2 \rfloor)\}, \quad (1)$$

$$P(k, m) = \{X(k - \lfloor l_p/2 \rfloor : k + \lfloor l_p/2 \rfloor, m)\}, \quad (2)$$

where stands for the median operator, $X$ is the magnitude of the input time-frequency representation, $k$ denotes the frequency bin index, $m$ represents the frame index, $l_h$ is the length of the harmonic median filter in time frames while $l_p$ is the length of the percussion median filter in frequency bins. Note that $l_p$ and $l_h$ should be odd values.

Next, a harmonic soft time-frequency mask $M_H$ through Wiener filtering is computed from the two previous filtered spectrograms,

$$M_H(k, m) = \frac{H^p(k, m)}{P^p(k, m) + H^p(k, m)}, \quad (3)$$

where $p$ refers to the exponent that is applied to every individual time-frequency filtered element. The mask $M_H$ isolates the harmonic content since it extracts the relative energy ratio of the harmonic components with respect to the entire energy of the input magnitude time-frequency representation. As a result, a harmonic-enhanced spectrogram $X_H$ is computed multiplying the harmonic mask $M_H$ and the input magnitude time-frequency representation $X$,

$$X_H(k, m) = M_H(k, m) \cdot X(k, m) \quad (4)$$

Because the predominant energy of snoring sounds is concentrated in the lower spectral range, preliminary results indicated a higher detection performance using CQT as an individual time-frequency representation compared to STFT and Mel. Consequently, the CQT was applied as the input time-frequency representation for the HPSS method, as depicted in Figure 2. In terms of the median filtering process, the length of the harmonic filter has been specifically set to $l_h = 33$ time frames (approximately 350 ms) to ensure that only harmonic events that satisfy the typical minimum snore duration are extracted. Focusing on the length $l_p$ of the percussive filter, it is variable in terms of the frequency bin index $k$ since the CQT does not provide a linear frequency scale. In this work, the percussive length $l_p$ has been adjusted to achieve a filter bandwidth approximately equal to one-sixteenth of the center frequency $f_k$ to be filtered. Finally, the harmonic soft mask $M_H$ obtained from HPSS applies

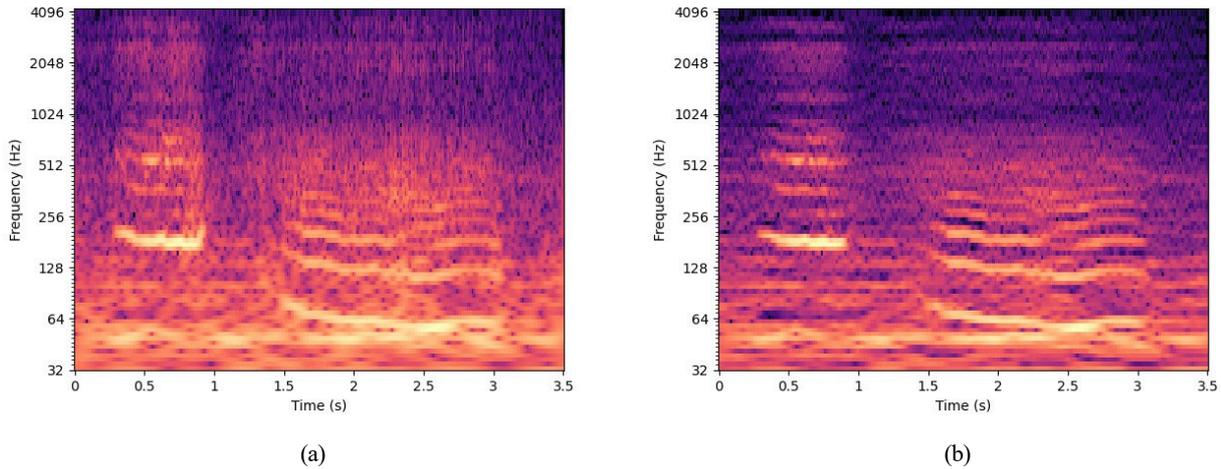

(a)                                                                 (b)

**Figure 3:** Time-frequency representation of a 3.5-second snore event. (a) CQT spectrogram. (b) Harmonic-enhanced spectrogram from HPSS.

Wiener filtering by means of $p = 2$, since it helps to emphasize the harmonic behavior of the snoring sounds. As an example, Figure 3 shows a comparison of the CQT spectrogram and the harmonic-enhanced spectrogram. It clearly observed the effectiveness of the harmonic-enhanced spectrogram in accentuating and improving harmonic features of the snoring sound during two-time intervals, from 0.3 to 1 second and from 1.5 to 3 seconds.

### 3.3. Stage 3. Neural network architecture

The baseline model for the neural network architecture is based on the approach described in [15], which is composed of three CNN layers and one LSTM layer. In addition to the baseline model, several standard architectures have been analyzed, specifically, VGG19 [69], MobileNet [70], and ResNet50 [71]. Table 2 presents a brief comparison among the state-of-the-art CNN architectures employed in this work.

To ensure methodological consistency with the method outlined in [15], a 5-fold cross-validation approach was adopted for training and testing the previous neural network architectures. Specifically, we implemented a leave-k-subjects-out cross-validation methodology, in which data from 27 patients were employed exclusively for testing, constituting 10% of the dataset. Simultaneously, the data from the remaining patients (242 subjects) was utilized as the training and validation set within a 5-fold cross-validation framework. To promote diversity in the training, validation, and test sets, we conducted the experiments iteratively 10 times, ensuring that each patient was included in the test set only once. Furthermore, this iterative process also minimizes the variance of the performance metrics. Considering the scenario with a constrained data set (see section 4.3), a transfer learning strategy was implemented for VGG19, MobileNet, and ResNet50 architectures. This approach was motivated by the demonstrated efficacy of transfer learning in addressing similar challenges associated with limited data [72, 73, 74, 75]. Specifically, for these neural network architectures, the convolutional layers were initialized using the pre-trained weights from the ImageNet dataset [76] and kept fixed throughout the training process. Subsequently, a trainable dense Softmax layer consisting of two units was appended after the non-trainable layers. Note that several recent works utilized transfer learning from specifically audio (mainly speech) architectures [77, 78]. However, rather than focusing purely on improving the results for the task of snore detection, in this paper, we aim to investigate the relevance of isolating the harmonic components inherent in snore sounds in order to increase the robustness of the detection methods against interferences.

|  | VGG19 | MobileNet | ResNet50 |
|---|---|---|---|
| Convolutional layers | 16 | 28 | 50 |
| Dense layers | 3 | - | - |
| Total layers | 19 | 28 | 50 |
| Total parameters | 20,025,410 | 3,230,914 | 23,591,810 |

**Table 2**
Comparative analysis of the layers and total parameters for VGG19, MobileNet, and ResNet50.

During the training phase, the adaptive moment estimation algorithm (Adam) was used as the optimizer, with a learning rate of $10^{-3}$. Besides, the binary cross-entropy loss function was employed, while the batch size was set to 64 and the number of epochs was set to 100. In order to prevent overfitting, an early stopping criterion was implemented during training, using the validation loss as the monitoring parameter. Specifically, the training was stopped if there was no improvement in the validation loss for 10 consecutive epochs, similarly as occurs in [79, 80].

## 4. Experimental results

In this section, we describe the metrics utilized to evaluate the proposed snore detection system and present the results obtained in two different scenarios: 1) using a large dataset of snoring and interfering sounds, and 2) using a reduced version of the dataset under a limited data learning scenario. It is noteworthy that the snoring detection task performed in this work focuses on identifying the audio signal class, aligning with the approach adopted in recent and referenced works on the subject [15, 29, 81].

### 4.1. Metrics

To evaluate the snoring detection performance of the proposed method, a set of metrics has been selected, in particular, accuracy ($Acc$), sensitivity ($Sen$), specificity ($Spe$), precision ($Prec$), score ($Sco$) and F1-score ($F1$). The mathematical formulations for quantifying each of these metrics are delineated herewith:

$$Acc = \frac{TP + TN}{TP + TN + FP + FN}, \quad (5)$$

$$Sen = \frac{TP}{TP + FN}, \quad (6)$$

$$Spe = \frac{TN}{TN + FP}, \quad (7)$$

$$Prec = \frac{TP}{TP + FP}, \quad (8)$$

$$Sco = \frac{Sen + Spe}{2}, \quad (9)$$

$$F1 = \frac{2 \cdot Prec \cdot Sen}{Prec + Sen}. \quad (10)$$

where $TP$ (True Positive) indicates the amount of snoring events classified as snoring, that is, the ability to correctly detect snoring sounds; $TN$ (True Negative) represents the number of interfering sounds that are correctly classified as interfering sounds, that is, the ability to correctly detect interfering sounds; $FP$ (False Positive) is the amount of interfering sounds classified as snoring; and $FN$ (False Negative) denotes the number of snoring events classified erroneously as interfering sounds.

### 4.2. Results using a large dataset of snoring and interfering sounds

In this scenario, from now on referred to as Scenario 1, the snore detection performance of the proposed method is evaluated using the complete training set from the database $D_T$.

Figure 4 shows the accuracy classification of each conventional neural network architecture when is fed by each STFT, Mel, CQT, and harmonic feature. Results indicate that the proposed method achieves the best performance with the baseline model and VGG19, while it provides comparable results to those obtained with the CQT when evaluated with ResNet50. Specifically, the proposed harmonic feature achieves average accuracy values of 96.5%, 93.3%, and 90.6% for the baseline model, ResNet50, and VGG19, respectively. Nonetheless, in the case of MobileNet, the CQT outperforms the proposed method, achieving an average accuracy improvement of 1%. Furthermore, the STFT consistently achieves better accuracy results compared to the Mel spectrogram. This insight suggest that the Mel spectrogram may not be the most appropriate time-frequency representation for accentuating the low-frequency characteristics associated with snoring. Regarding the compared architectures, the baseline model provides the most accurate results, independently of the used input representation. ResNet50 provided the best performance among the compared CNN architectures, followed by MobileNet and VGG19. Note that the performance of MobileNet drops substantially when using STFT or the Mel-based input representations. Therefore, it seems that residual neural networks provide the most optimal performance for snore detection compared to the standard CNNs employed in this study as indicated in [41].

| Comparison | Wilcoxon signed-rank test (p-value) | Significantly better |
|---|---|---|
| Baseline | | |
| CQT vs STFT | 0.000130 | yes |
| CQT vs Mel | 1.24e-14 | yes |
| Proposed method vs CQT | 0.000915 | yes |
| VGG19 | | |
| CQT vs STFT | 1.78e-15 | yes |
| CQT vs Mel | 3.55e-15 | yes |
| Proposed method vs CQT | 0.000109 | yes |
| MobileNet | | |
| CQT vs STFT | 1.78e-15 | yes |
| CQT vs Mel | 1.78e-15 | yes |
| Proposed method vs CQT | 1.49e-08 | no |
| ResNet50 | | |
| CQT vs STFT | 9.75e-08 | yes |
| CQT vs Mel | 1.78e-15 | yes |
| Proposed method vs CQT | 0.758304 | no |

**Table 3**
Wilcoxon signed-rank test conducted on the accuracy data obtained in Scenario 1, using a significance level $\alpha = 0.05$.

To evaluate the statistical significance of the influence of different features (i.e. spectrograms from CQT, STFT, Mel, and the proposed feature) on snoring detection within a substantial dataset, the Wilcoxon signed-rank test [82] was employed. This analysis was conducted for all the evaluated neural network architectures. In this statistical test, the null hypothesis $H_0$ states that there is no significant difference between the results from different features. The $p$-value with a predefined significance level $\alpha < 0.05$ indicates statistically significant and $H_0$ could be rejected. The statistical results, shown in Table 3, indicate that the CQT spectrogram significantly improves the accuracy performance compared to the STFT and Mel spectrogram. Furthermore, the proposed spectrogram, based on the combination of the cascaded CQT with HPSS, yields a significant improvement compared to the individual use of the CQT spectrogram when both are evaluated with the baseline model and VGG19. However, for MobileNet and ResNet50, Table 3 demonstrates the absence of a significant improvement of the proposed method with respect to CQT. Finally, it should be noted that, although the obtained $p$-value for the comparison is lower than the significance level $\alpha$ in the case of MobileNet, it is the CQT feature itself that is responsible for this improvement, rather than the proposed method.

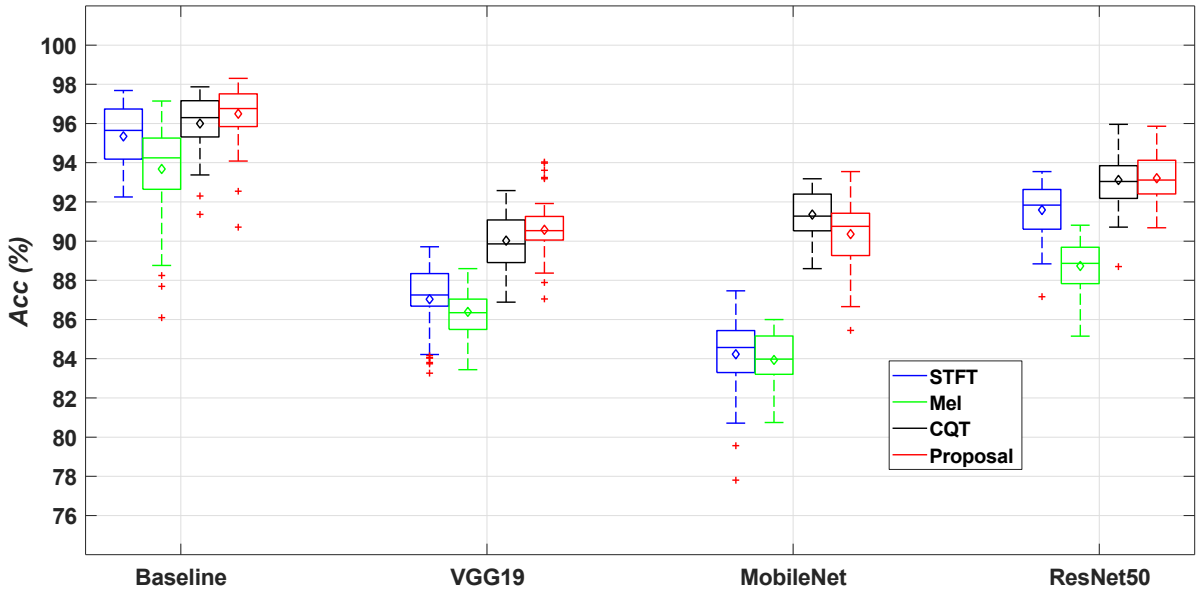

**Figure 4:** Accuracy performance of the proposed method, using as feature extraction the harmonic extracted content (red) compared to STFT (blue), Mel (green), and CQT (black), feeding into several standard neural network architectures, namely the baseline model [15], VGG19, MobileNet and ResNet50 in Scenario 1. The data used for analysis was obtained from a 5-fold cross-validation process performed in 10 experiments, with each box in the resulting plots representing 50 data points. The first and third quartiles are shown by the lower and upper lines of each box, respectively, and the median value is depicted as the line located within the box. The diamond shape in the center of each box illustrates the average value. Red crosses represent the outliers, which are values more than 1.5 times the interquartile range away from the lower or upper end of the box. The range of the remaining data is illustrated by the dashed lines that extend above and below each box.

Although the results obtained by the proposed HPSS-based feature are promising when using a large dataset of training sounds, in this paper, we put a focus on the interesting behaviour of the proposed feature when dealing with limited training data (Scenario 2) in Section 4.3. Consequently, to keep the presentation of the paper compact, further analysis of the results of Scenario 1 (i.e. large training dataset) using additional metrics from Section 4.1 as well as the accuracy, loss and ROC curves are presented in Appendix A.

### 4.3. Results using a small dataset of snoring and interfering sounds

In this scenario, from now on referred to as Scenario 2, a new training database that consists of approximately 1% of the snoring and interfering sounds from the training set used in Scenario 1 was employed to evaluate the snoring detection performance of the proposed method in a limited data learning context. The sounds were selected using a random procedure aimed to mitigate any reliance on specific audio file selections and obtain more robust and reliable results. Specifically, this new training database $D_R$ was composed of approximately 90 snoring sounds and 60 interfering sounds for each one of the 10 experiments conducted based on 5-fold cross-validation. To validate the results, the test set is the same as the one used in Scenario 1.

Accuracy results in Figure 5 indicate that the harmonic feature of the proposed method consistently outperforms the performance of the STFT, Mel, and CQT features for all neural network architectures evaluated in this learning scenario. These average accuracy values achieved by the proposed method range from 78.8% to 82.8%, emphasizing its stability and robustness across different learning models employed in the context of snoring detection. Focusing on the type of neural network architecture evaluated, it is noticeable that ResNet50 consistently achieves the highest accuracy results across all evaluated spectrograms. This finding suggests that the residual connections in ResNet50 contribute to a more effective characterization of snoring, particularly when dealing with limited available data. Nevertheless, two distinct behaviors can be observed. The first trend pertains to the baseline model and MobileNet, where both the CQT and the proposed method demonstrate significantly superior performance, showing at least an improvement of 9% of the proposed method compared to the STFT and Mel spectrogram. However, even in the worst case, the proposed method shows an improvement of over 6%. This suggests that the proposed harmonic feature provides more reliable learning of the most relevant time-frequency characteristics of the snoring sounds since a lower interquartile range can be observed in most of the neural network architectures evaluated.

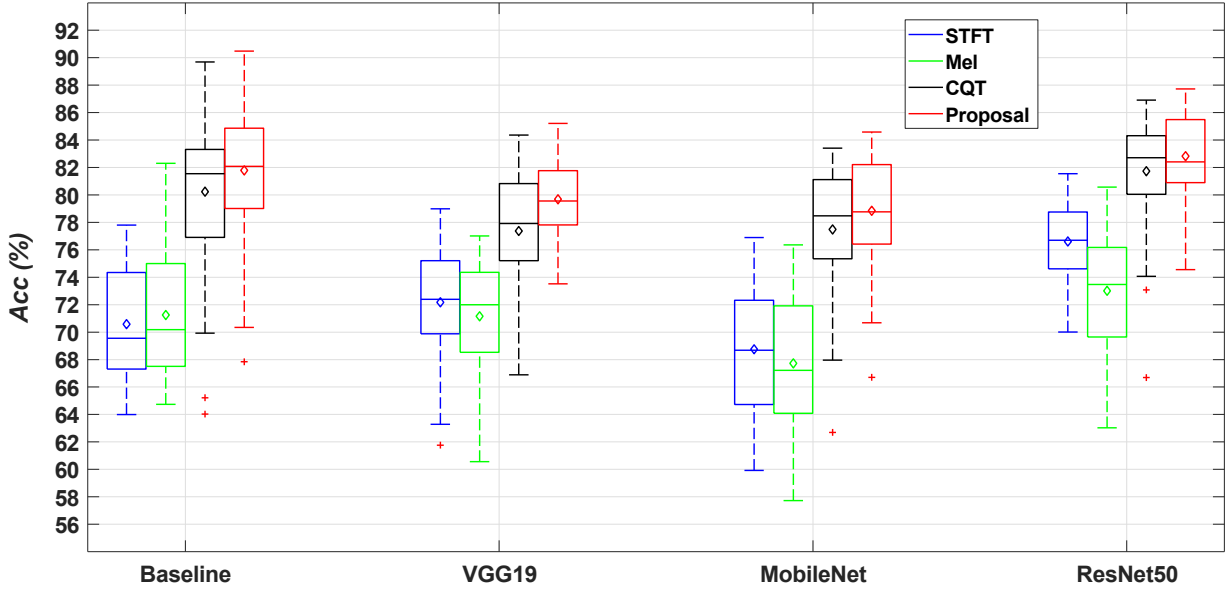

**Figure 5:** Accuracy performance of the proposed method, using as feature extraction the harmonic extracted content (red) compared to STFT (blue), Mel (green) and CQT (black), feeding into several standard neural network architectures, namely the baseline model [15], VGG19, MobileNet and ResNet50 in Scenario 2. The data used for analysis was obtained from a 5-fold cross-validation process performed in 10 experiments, with each box in the resulting plots representing 50 data points. The first and third quartiles are shown by the lower and upper lines of each box, respectively, and the median value is depicted as the line located within the box. The diamond shape in the center of each box illustrates the average value. Red crosses represent the outliers, which are values more than 1.5 times the interquartile range away from the lower or upper end of the box. The range of the remaining data is illustrated by the dashed lines that extend above and below each box.

By comparing Figure 4 and Figure 5, it can be observed the higher variability in the accuracy results obtained in Scenario 2. This fact highlights that, under data-limited conditions, snoring detection performance is strongly affected by the particular snoring characteristics inherent to each subject. Besides, it can also be observed that, unlike in Scenario 1, the proposed method shows significant superiority over the CQT when it is trained on a small dataset $D_R$, independently of the model. This fact emphasizes the effectiveness of the proposed harmonic feature in such limited data scenarios. Specifically, the proposed method demonstrates an accuracy improvement of 1.6% with the baseline model, 2.3% with VGG19, 1.3% with MobileNet, and 1.1% with ResNet50, compared to the CQT. Figure 6 highlights that the proposed method consistently outperforms the other approaches across all evaluated metrics. When comparing the results for sensitivity (Figure 6a) and specificity (Figure 6b), it is evident that the architectures show superior performance in snoring detection compared to non-snoring detection. This is corroborated by the higher sensitivity values, indicative of a greater ability to accurately identify snoring. In contrast to Scenario 1, the proposed method surpasses the other approaches in terms of sensitivity but also in terms of specificity and precision (Figure 6c). As a result, these outcomes demonstrate that in a limited data learning scenario, where only a few examples are available for learning, the proposed signal processing technique HPSS applied before the neural network architecture enhances the ability to distinguish more reliably between snoring and other sounds by focusing on the harmonic content that is often present in most snore sounds, as it is also confirmed in terms of score (Figure 6d) and F1-score (Figure 6e), improving the learning capabilities for all neural network architectures evaluated in this study.

Table 4 presents the outcomes obtained with the Wilcoxon signed-rank test [82] in Scenario 2. In this scenario, the statistical results also confirm that the CQT spectrogram enhances the accuracy performance compared to STFT and Mel spectrogram. Nonetheless, the main difference compared to Scenario 1 lies in the considerable enhancement showcased by the proposed harmonic spectrogram when compared to using the CQT spectrogram individually, regardless of the neural network architecture.

In order to observe the behavior of the accuracy models, loss models, and ROC curves in Scenario 2, Figure 7 illustrates the results obtained for a case evaluated with ResNet50, chosen due to its superior performance in this scenario. Comparing the accuracy (Figure 7j) and loss (Figure 7k) models obtained using the proposed harmonic feature with the other features, it becomes evident that the feature based on harmonic content outperforms the rest of features, demonstrating better learning between the training and validation results shown in smoother curves with fewer abrupt transitions. This fact implies a more stable overall performance of the proposed method. In terms of ROC curves, the proposed feature also outperforms the others, achieving an AUC of 0.890 compared to 0.877 (CQT), 0.852 (STFT), and 0.843 (Mel). These AUC values confirm that the proposed method offers the best discrimination ability to distinguish snoring from non-snoring sounds in a limited data learning context.

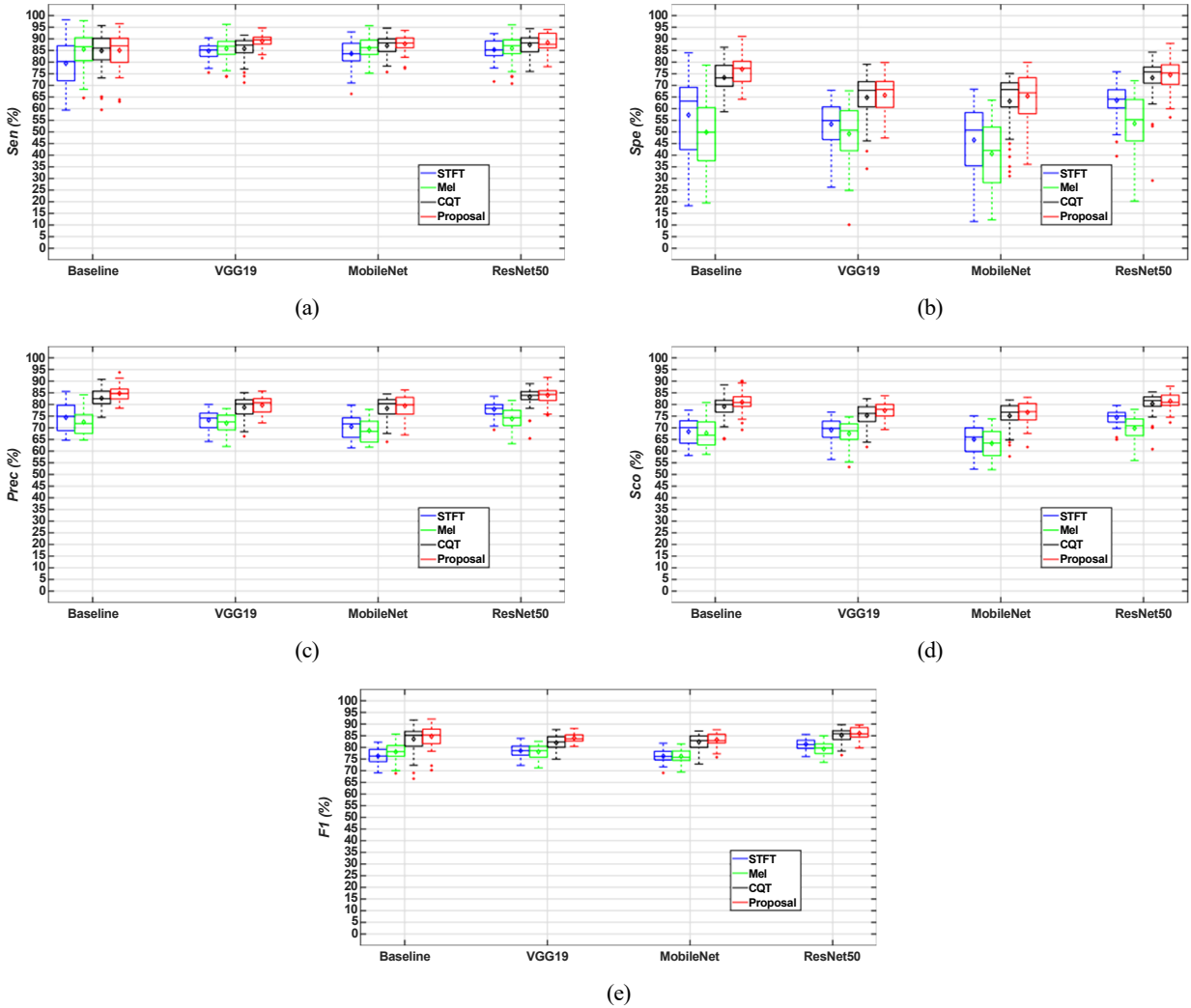

**Figure 6:** Sensitivity (a), specificity (b), precision (c), score (d) and F1-score (e) performance of the proposed method, using as feature extraction the harmonic extracted content (red) compared to STFT (blue), Mel (green) and CQT (black), feeding into several standard neural network architectures, namely the baseline model [15], VGG19, MobileNet and ResNet50 in Scenario 2. The data used for analysis was obtained from a 5-fold cross-validation process performed in 10 experiments, with each box in the resulting plots representing 50 data points. The first and third quartiles are shown by the lower and upper lines of each box, respectively, and the median value is depicted as the line located within the box. The diamond shape in the center of each box illustrates the average value. Red crosses represent the outliers, which are values more than 1.5 times the interquartile range away from the lower or upper end of the box. The range of the remaining data is illustrated by the dashed lines that extend above and below each box.

The previous results obtained in Scenario 2 manifest the critical relevance of employing the proposed harmonic feature to accurately detect snoring sound in limited data contexts. Specifically, this insight was derived using as a training data a 1% of the complete training dataset (approximately 90 snoring sounds and 60 interfering sounds). However, in order to determine how the performance of the proposed method evolves as the volume of training data increases, different training databases were composed using 1% increments in the amount of data with respect to the training dataset employed in Scenario 2 until reaching the 10%.

Figure 8 represents the average accuracy results obtained with the different training data sizes. Based on this figure, it can be observed that, when the training data increase to the 2% (approximately 180 snoring sounds and 120 interfering sounds), the proposed harmonic feature still achieves the best performance when evaluated with the baseline model and VGG19, achieving an average accuracy improvement of 1.2% and 0.9% over the CQT, respectively. However, unlike Scenario 2, the best performance is not achieved with ResNet50, but when the proposed feature is evaluated with the baseline model, reaching an average accuracy value of 88.0%. On the other hand, the CQT slightly outperforms the proposed harmonic feature for MobileNet and ResNet50, providing respective improvements of 0.4% and 0.2% in mean accuracy.

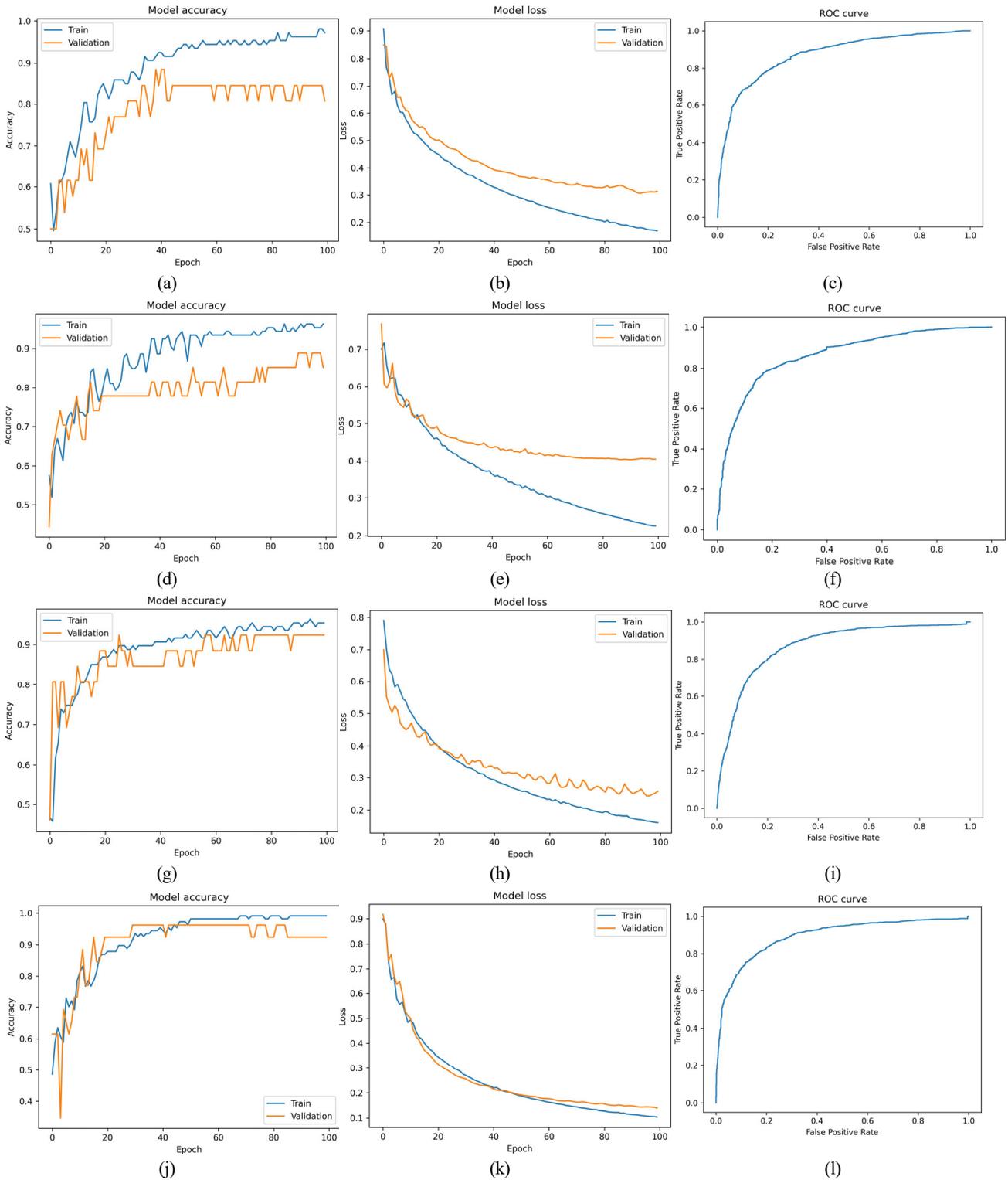

**Figure 7:** Model accuracy, model loss, and ROC curves when the STFT (a, b, c), Mel (d, e, f), CQT (g, h, i) and the proposed method (j, k, l) are evaluated with ResNet50 in the Scenario 2.

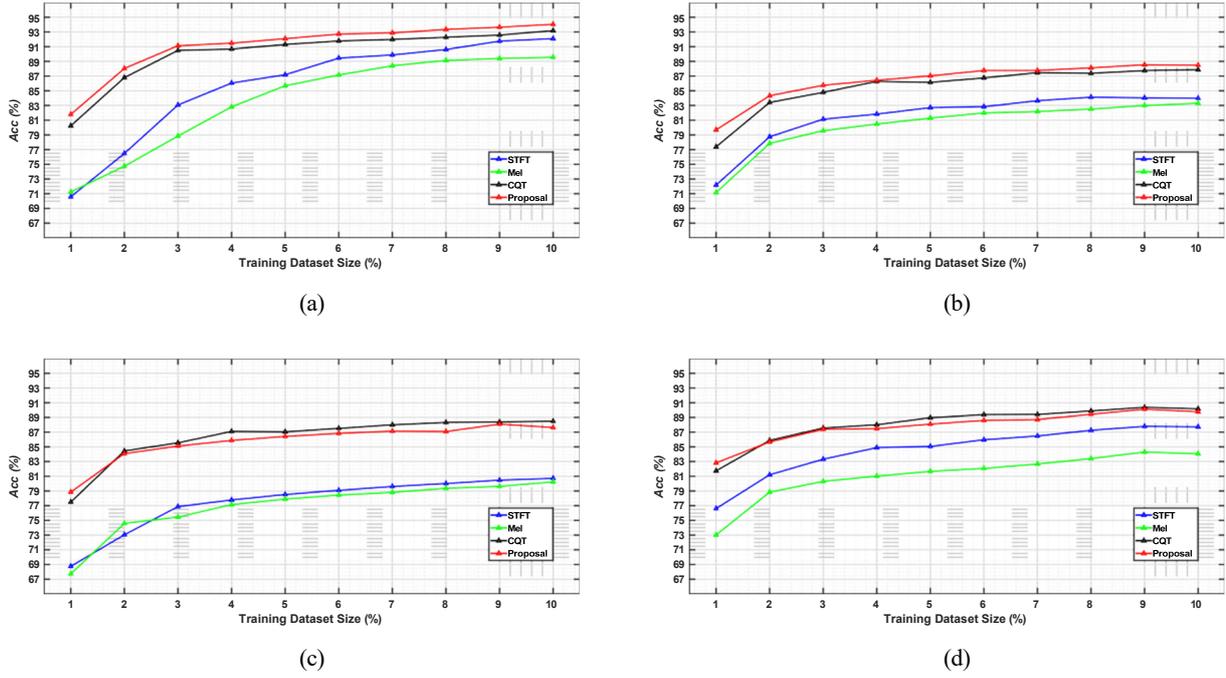

**Figure 8:** Average accuracy performance of the proposed method, using as feature extraction the harmonic extracted content (red) compared to STFT (blue), Mel (green) and CQT (black), feeding into several standard neural network architectures, namely the baseline model [15] (a), VGG19 (b), MobileNet (c) and ResNet50 (d) using training data sizes ranging from 1% to 10%.

| Comparison | Wilcoxon signed-rank test ($p$-value) | Significantly better |
|---|---|---|
| **Baseline** | | |
| CQT  vs  STFT | 7.94e-13 | yes |
| CQT  vs  Mel | 2.43e-13 | yes |
| Proposed method vs CQT | 0.010424 | yes |
| **VGG19** | | |
| CQT  vs  STFT | 1.26e-08 | yes |
| CQT  vs  Mel | 2.26e-10 | yes |
| Proposed method vs CQT | 2.55e-05 | yes |
| **MobileNet** | | |
| CQT  vs  STFT | 2.49e-14 | yes |
| CQT  vs  Mel | 1.34e-09 | yes |
| Proposed method vs CQT | 0.008849 | yes |
| **ResNet50** | | |
| CQT  vs  STFT | 1.35e-11 | yes |
| CQT  vs  Mel | 5.33e-15 | yes |
| Proposed method vs CQT | 0.022378 | yes |

**Table 4**
The Wilcoxon signed-rank test conducted on the accuracy data obtained in Scenario 2, using a significance level $\alpha = 0.05$.

By comparing the results obtained using the 2% of the training database with the results achieved with the remaining higher training data sizes, it can be observed a similar behaviour to the one observed when using the 2%. Furthermore, it is also noticeable the accuracy improvement showcased with all the input time-frequency representations when the training data increase from 1% to 2%. Specifically, the mean accuracy improvement with respect to Scenario 2 ranges from 2.8% to 6.2% in the case of the proposed harmonic feature.

## 5. Discussion

In this work, the learning performance of CNN models in a scenario with limited data for snore detection is improved. This enhancement is achieved by employing harmonic spectrograms derived from HPSS, instead of conventional time-frequency representations. A comprehensive assessment to evaluate the performance of the proposed method in detecting snoring versus various types of non-snoring sounds, commonly encountered in household and hospital settings, has been conducted. Specifically, snoring sounds were randomly mixed with clinical ambient sounds, household noises, room sounds, and cough sounds using different signal-to-noise ratios. The aim was to replicate realistic acoustic scenarios and determine how robust the proposed method could identify snoring in several real-world situations.

In Scenario 1, the proposed method provides superior or comparable results to those obtained with the CQT, except in the case of MobileNet, where the CQT clearly outperforms the proposed HPSS-based feature. This unexpected behaviour of CQT, revealing that the harmonic content extracted from the CQT spectrogram is less crucial than the raw data from the same spectrogram in this scenario, could be explained by the lower number of parameters of MobileNet compared to VGG19 and ResNet50, as shown in Table 2. Nevertheless, comparing the performance of the different detection metrics provided by the evaluated architectures, it is concluded that in a scenario where there is considerable data availability, the baseline model, based on the CNN-LSTM model, fed with the proposed harmonic content improves the extraction of reliable information that faithfully represents the occurrence of a snore that conventional CNNs fail to capture.

Furthermore, the proposed method exhibits a highly promising performance in the context of limited data learning (Scenario 2). The obtained results demonstrate superior detection capabilities across all metrics and CNN architectures, significantly surpassing the performance of alternative spectrograms such as STFT, Mel, and CQT. Additionally, the proposed method effectively mitigates the presence of outliers, as illustrated in Figure 5. Results suggest that incorporating harmonic content significantly enhances the robustness of snore detection, thereby confirming our initial hypothesis that features based on harmonic content improve the reliability of limited data learning in snore detection. Specifically, the performance obtained in this scenario indicates that the strength of the proposed method lies in its ability to accurately discriminate between snoring and non-snoring sounds, effectively capturing the most distinctive spectral structure predominantly present in most snore sounds as shown in Figure 6b. Moreover, although the contribution of the harmonic spectrogram may not be immediately apparent, it plays a crucial role in snoring detection. While it only leads to a slight improvement in sensitivity, as depicted in Figure 6a, it substantially enhances precision, as depicted in Figure 6c. Consequently, this improvement in precision results in a significantly higher proportion of correctly classified snoring sounds. These findings suggest that the harmonic feature, derived from the HPSS-based harmonic spectrogram, is able to capture the essential spectral content that characterizes snoring in most instances.

To investigate the impact of the training data quantity on the proposed method performance, additional experiments with incremental training datasets were conducted in order to explore the performance modulation as the training data volume increased. The different results obtained suggest that the proposed harmonic feature is crucial in contexts with extremely limited data (Scenario 2). However, under conditions in which the data availability is not so limited (≥ 2%), the proposed method achieves superior or comparable results to those obtained by CQT, although its improvement depends on the classification architecture employed.

## 6. Conclusions and Future work

Empirical observations have revealed that a significant number of snoring sounds exhibit a strongly harmonic behaviour. This study explores the potential benefits of leveraging these harmonic features in scenarios with limited training data. This work proposes incorporating the harmonic behavior commonly found in monaural snoring sounds to enhance the learning process of snore detection. In contrast to conventional reliance on traditional time-frequency transforms, such as STFT spectrogram, CQT spectrogram, and Mel-scaled spectrogram, our approach focuses on incorporating the extracted harmonic content through the application of harmonic/percussive sound source separation (HPSS) technique. By doing so, we aim to improve the learning process within conventional neural network architectures, enabling more effective snore detection even with limited training data, and emphasizing the potential of the harmonic feature and HPSS as valuable tools in the domain of limited data learning. A comprehensive evaluation has been performed to analyze the effects of different data-feeding techniques on snoring detection using conventional CNN architectures. The evaluation involved widely recognized repositories of snoring sounds, non-snoring sounds, and mixed signals with varying signal-to-noise ratios. Furthermore, we compared our approach with several deep learning architectures. The results indicated that our method performed competitively when a large number of samples of both snoring and non-snoring sounds were available. Moreover, our approach exhibited statistically significant improvements and more stable behaviour in the learning curves during limited data learning compared to other input time-frequency transforms across all tested CNN models.

Future work will focus on two main directions. First, we will work in the identification of the severity of Obstructive Sleep Apnea Syndrome (OSAS) cases based on snoring patterns. Second, we will explore the use of transfer learning strategies from specifically designed audio architectures, to enhance the interpretability of neural network-based snoring detection models since it holds critical importance in medical applications.

### A. Additional results using a large dataset of snoring and interfering sounds

In this section, we provide further analysis of the results of scenario 1 using additional metrics from Section 4.1 as well as an analysis of the accuracy, loss and ROC curves.

First of all, Figure 9 presents the outcomes achieved for the remaining evaluated metrics. Independently of the utilized input representation, the baseline model obtains the best score (Figure 9d) and F1-score (Figure 9e) results. In fact, the performance of the compared input representations is similar in terms of specificity (Figure 9b) and precision (Figure 9c) while, in the case of sensitivity, slightly worse results are obtained (Figure 9a) and the best performance is observed using the proposed harmonic representation ($Sen$ = 96.4%). These results imply that the system tends to classify interfering sounds more accurately than snoring sounds. Regarding the performance of the other compared architectures, higher variability in their results can be observed depending on the chosen input representation. In general, CQT and the proposed harmonic feature provide the best performance for all the metrics and methods. In particular, similar score performance is observed for VGG19 and ResNet50 while, in the case of MobileNet, the CQT ($Sco$ = 91.2%) outperforms the proposed harmonic feature ($Sco$ = 90.2%). One possible reason for this similar detection level seems the imbalance of the dataset among the different classes of sounds, as was previously mentioned in Section 2. However, even though the performance results provided by the harmonic content are promising and competitive, it does not demonstrate a significant improvement compared to using the raw data directly from CQT in this scenario in which a large amount of training data is available.

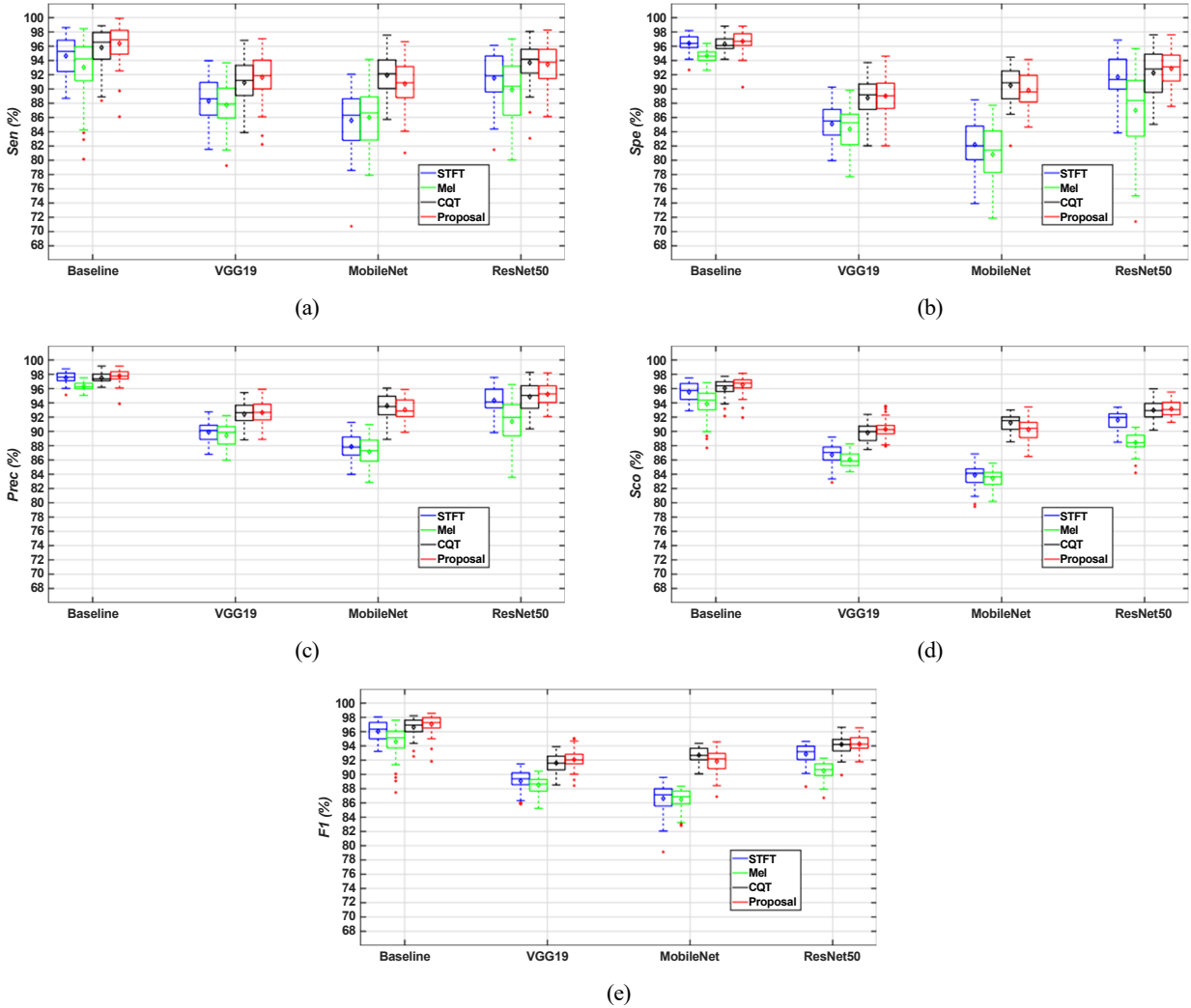

**Figure 9:** Sensitivity (a), specificity (b), precision (c), score (d) and F1-score (e) performance of the proposed method, using as feature extraction the harmonic extracted content (red) compared to STFT (blue), Mel (green) and CQT (black), feeding into several standard neural network architectures, namely the baseline model [15], VGG19, MobileNet and ResNet50 in Scenario 1. The data used for analysis was obtained from a 5-fold cross-validation process performed in 10 experiments, with each box in the resulting plots representing 50 data points. The first and third quartiles are shown by the lower and upper lines of each box, respectively, and the median value is depicted as the line located within the box. The diamond shape in the center of each box illustrates the average value. Red crosses represent the outliers, which are values more than 1.5 times the interquartile range away from the lower or upper end of the box. The range of the remaining data is illustrated by the dashed lines that extend above and below each box.

As an example, Figure 10 illustrates the accuracy models (first column), loss models (second column), and ROC curves (third column) obtained for a random case assessed with the baseline model in Scenario 1. Comparing the obtained accuracy and loss models, a notable similarity between the results obtained from CQT and the proposed harmonic feature can be clearly observed. Additionally, both spectrograms exhibit better performance compared to STFT and Mel spectrogram, which also offer similar results between them. This superior performance is indicated by the lower number of epochs required for CQT and the proposed method, demonstrating faster convergence in the process of learning the distinctive features of snoring sounds. In terms of ROC curves, the proposed feature and CQT also demonstrate superior performance compared to STFT and Mel spectrogram, as both achieve AUC (Area Under the ROC Curve) values of 0.993 and 0.992, respectively, while the Mel spectrogram yields 0.987 and the STFT obtains 0.983.

---

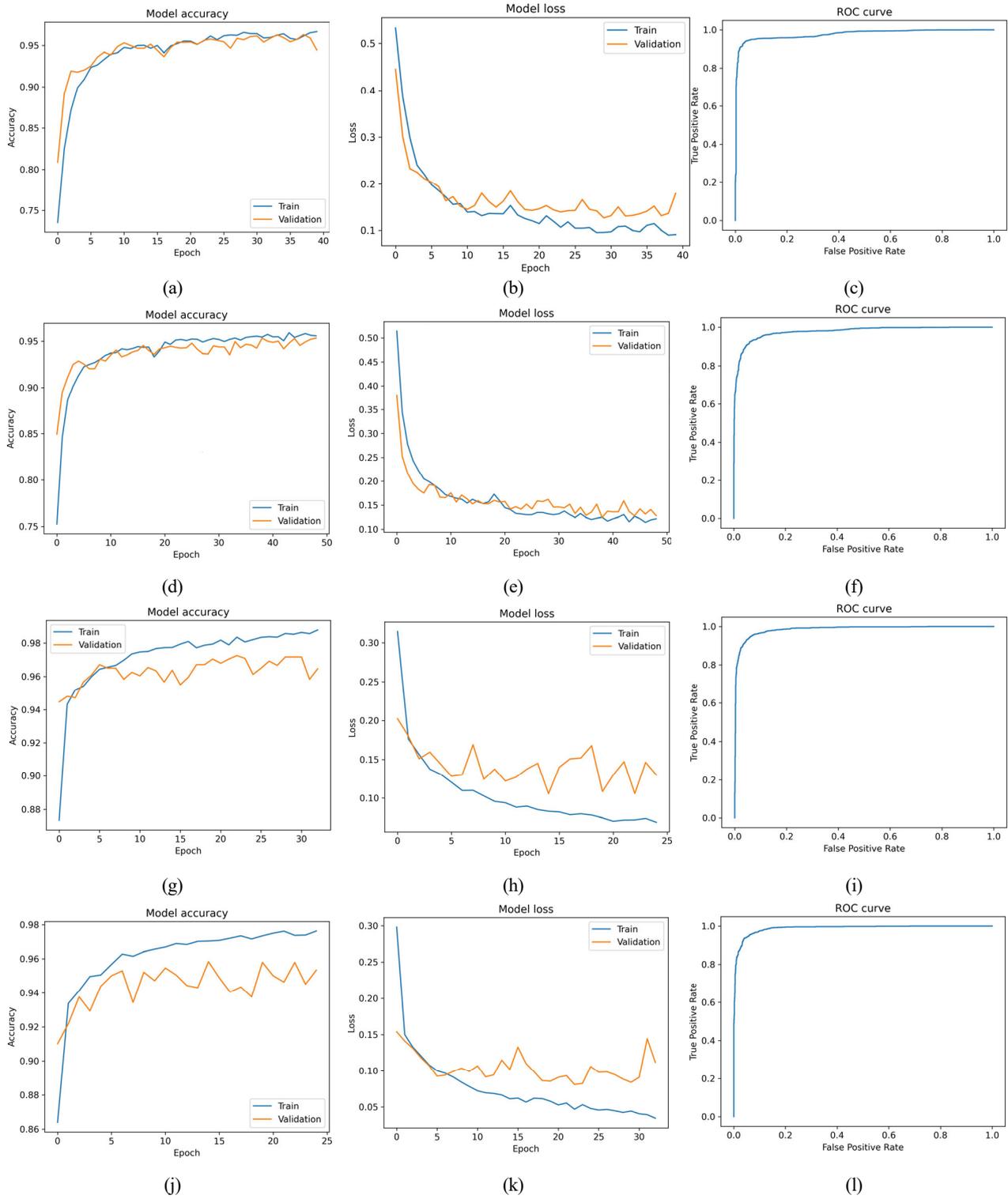

**Figure 10:** Model accuracy, model loss, and ROC curves when the STFT (a, b, c), Mel (d, e, f), CQT (g, h, i) and the proposed method (j, k, l) are evaluated with the baseline model [15] in the Scenario 1.